\documentclass[12pt,a4paper]{article}
\pdfoutput=1
\usepackage{amsmath}
\usepackage{amsthm,amssymb}
\usepackage{rotating}
\usepackage{pdflscape}
\pdfoutput=1 
\numberwithin{equation}{section}
\usepackage{graphicx}

\title{\Large \bf    Seasonal Entropy, Diversity and Inequality Measures of Submitted and Accepted   Papers Distributions In  Peer-Reviewed  Journals 
 }

\author{  \large \bf  Marcel    Ausloos$^{1,2,3,^*}$, Olgica Nedic$^{4}$, Aleksandar Dekanski$^{5}$
  \\ \\$^{1}$  School of Business,  College of Social Sciences, Arts, and Humanities, \\University of  Leicester, Leicester,   LE2  1RQ, United Kingdom \\  email:  ma683@le.ac.uk
 \\$^{2}$   Department of Statistics and Econometrics, \\Bucharest University of Economic Studies, \\ Calea Dorobantilor 15-17, Bucharest, 010552 Sector 1, Romania \\
  email: marcel.ausloos@ase.ro
  \\$^{3}$   Group of Researchers for Applications of Physics \\in Economy and Sociology (GRAPES), \\Rue de la belle jardini\`ere, 483, B-4031 Angleur, Li\`ege, Belgium \\  
 email: marcel.ausloos@ulg.ac.be 
 \\$^{4}$  Institute for the Application of Nuclear Energy (INEP),\\
University of Belgrade,  Banatska 31b, Belgrade-Zemun, Serbia 
  \\$^{5}$  Institute of Chemistry, Technology and Metallurgy, 
\\Department of Electrochemistry,\\ University of Belgrade,   Njegoseva12,  Belgrade, Serbia  }

\begin{document}
\maketitle  
\newpage
\begin{abstract} 

This paper presents a novel method for finding features in the analysis of variable distributions stemming from   time series. We apply the methodology to the case  of submitted and accepted papers in peer-reviewed journals.  We  provide  a  comparative study  of editorial decisions for papers submitted to  two peer-reviewed journals:  the Journal of the Serbian  Chemical  Society ({\it JSCS}) and this MDPI  {\it Entropy} journal. We cover three recent years for which the fate of submitted papers, about 600 papers to  {\it JSCS}  and 2500 to {\it Entropy}, is completely determined. 

 Instead of comparing the number distributions of these papers as a function of time with respect to a uniform distribution,  we  analyze the  relevant probabilities, from which we derive the information entropy.  It is argued that such probabilities  are indeed more relevant for authors than the actual number of submissions. 
 
We tie this entropy analysis to the so called diversity of the variable distributions. Furthermore, we emphasize the  correspondence between the entropy and the diversity with inequality measures, like the   Herfindahl-Hirschman index and  the Theil index, itself being  in the class of entropy measures;  the Gini coefficient which also measures  the diversity in ranking is calculated for  further discussion.

In this sample,  the seasonal aspects of the peer review process  are outlined. It is found that  the use of such indices, non linear transformations of the data distributions,  allow to distinguish features and evolutions of peer review process as a function of time as well as  comparing non-uniformity of  distributions. 
Furthermore, $t-$ and $z-$ statistical tests are applied in order to measure the significance ($p-$ level) of the findings, i.e. whether
papers are more likely to be accepted   if they are submitted during  a few specific months or "season"; the predictability strength depends on the journal.
 

\end{abstract}


 \newpage
\section{Introduction}\label{introduction}
 
 Authors, who submit (by their own assumption) high quality papers  to scholarly journals, are interested to know if there are  factors which may increase the probability that their papers   be accepted. One of such factors may be related to the month of submission, or to the day of submission, as recently discussed  \cite{SCIMHerteliu}.    Indeed,  authors might wonder about editors and reviewers overload at some time of the year. Moreover, the number of submitted papers is  relevant for editors and publishers handling machines to the point that  artificial intelligence     can  be useful  for helping journal editors  
 \cite{r7,r8}.  More generally,
 informetrics and bibliometrics are also interested in the manuscript submission timing especially in the light of an enormous increase in the number of electronic journals.
 
  From  the author point of view, the rejection is often  frustrating, be it due as a "editor desk rejection" or following a review process. One has sometimes  explained a high editor desk rejection rate due to  an entrance barrier editor load effect \cite{r10}.  Thus, it is of interest  to observe whether there is a high probability of submission during specific months or seasons. In fact, the  non uniform  submission has already been studied. However, the   acceptance distribution,  during  a   year, i.e. a "monthly bias", is  rarely studied, because of publisher secrecy.   Search engines do not provide any information at all on the timing of rejected papers.

 Interestingly, recently, Boja et al. \cite{SCIMHerteliu} examined a large database on journals with high impact factor and reported that  a {\it day of the week} correlation effect  occurs between  "when a paper is submitted to a peer-reviewed journal  (and) whether that paper is accepted".  However, bis, there was no study of rejected papers. - because of a lack of data Thus, one may wonder if beside a {\it "day of the week"} effect,  there is some "{\it seasonal"} effect. One may indeed imagine that researchers in academic surroundings do not have a constant occupation rate, due to  teaching classes, holidays, congresses, and even budgetary conditions. Researchers have only specific times during the academic year for producing research papers.

 From the {\it "seasonal effect"} point view,   Shalvi et al.  \cite{r11} 
 found a discrepancy in the pattern of "submission-per-month" and "acceptance-per-month" for  Psychological Science  ($PS$), - but not for Social Psychology Bulletin ($PSPB$).  
  Summer months inspired authors to submit more papers to  $PS$, but the subsequent acceptance was not related to the effect of seasonal bias    (based on  a $\chi^2_{(11)}$ test for percentages); on the other hand, a very low rate of acceptance was recorded for manuscripts sent in November or December. The number of submissions to $PSPB$, on the contrary, was the greatest during winter months, followed by a reduced "production" in April; however, the rate of the acceptance was the highest for papers submitted in the period [Aug.-Sept.-Oct.].  
Moreover, a significant “acceptance success dip” was noted for submissions made in winter months. One of the main reasons for such differences between journals was  conjectured to lie in different rejection policies; some journals employ desk rejection, whereas others do not.
 
 Schreiber  \cite{r10}  analysed the acceptance rate in a journal, Europhysics Letters ($EPL$), for a period of 12 years and found that the rate of manuscript submission exceeded the rate of their acceptance. The data revealed (Table 2 in  \cite{r10}) that there is a maximum in the number of submissions in July, defined as a 10 \% increase compared to the annual mean, together with a minimum in Feb., even taking into account the "smaller length" of this month. He concluded that significant fluctuations exist between months. The acceptance rate  was ranging from 45\% to 55\%; the highest acceptance rate was seen in July and the lowest in January,  in the most recent years.
 
Recently, Ausloos et al. \cite{SCIM19MAONAD}   studied  submission and also  subsequent acceptance data for two journals\footnote{a specialized (chemistry) scientific journal and  a multidisciplinary journal, respectively },  Journal of the Serbian Chemical Society  ({\it JSCS})\footnote{$http://shd.org.rs/JSCS/$.} and {\it Entropy}\footnote{ $http://www.mdpi.com/journal/entropy$. }, each over a 3 year time interval. The authors find that fluctuations, expectedly, occur:  the number of submissions for $JSCS$ is the greatest in July and Sept. and the smallest in May and Dec. The highest rate of paper submission for $Entropy$ was noted in Oct. and Dec. and the lowest in Aug. Concerning acceptance for $JSCS$, the proportion of accepted/submitted manuscripts is the greatest in Jan. and Oct. 
Concerning acceptance  for $Entropy$,  the number of papers steadily increase from January  till  a peak in May,  followed by a marked dip during summer time, before reaching a peak in October of the order of the May peak.

Concerning the number of submitted manuscripts, it was observed that the acceptance rate in $JSCS$ was  the highest if papers were submitted in January and February; it was  significantly low if the submission occurred in December. In the case of $Entropy$, the highest rejection rate was for papers submitted in December and March, thus with a January-February peak; the lowest acceptance rate was  for manuscripts submitted in June or December; the highest rate being for those sent in spring months, February to May. One recognizes  a journal seasonal shift\footnote{We adapt  the word "seasonal"; even though changes in seasons occur on the 21st of various months, we approximate the season transition as occurring on the next  1st day of the following month.} of the features. 
  

   
Here, we propose another line of approach in order to study the submission, acceptance, and rejection (number and rate) diversity,  based on probabilities, with emphasis on the conditional probabilities, thereafter measuring the entropy and other characteristics, of the distributions.  Indeed, the entropy is a measure of disorder, and one of several ways to measure diversity. 
Researchers have their own preference \cite{Marhuendaetal2005,AlizadehNoughabi2017} in measuring diversity. Here below, we practically adapt the classical measure of diversity, as used in ecology, but other cases of interest pertaining to information science \cite{Rousseau92,Loet11} can be mentioned.
 


Let us recall that the  general equation of diversity is often written in the form  \cite{Hill,Jost}
\begin{equation}
\label{diversityeq}
{^q}D =  \large[\sum_{i=1}^N p_{i}^q  ]\large^{1/(1-q)}
\end{equation}
  in which  $ p_i=[z_i/\sum_i z_i]$, and $z_i$ the measured variable. 
  For $q=1$,  ${^q}D$ reduces to the exponential of the Shannon entropy \cite{shannona,shannonb}
 \begin{equation}\label{diversity1}
 ^1D = exp[-\sum_{i=1}^N  p_i \;  ln(p_i)],
 \end{equation}
 to which we will only stick here.
 
Several  inequality  measures  are  commonly  used  in  the  literature:  
  in the class of entropy related measures, one finds the  exponential entropy  \cite{Campbell66}, which measures the extent of a distribution, and the Theil index \cite{Theil1967} which emerges as the most popular one \cite{Beirlanetal01,OanceaPirjol19}, beside the Hirschman-Herfindahl  index  \cite{Hirschman}, measuring "concentrations".
 "Finally", upon ranking according to their size the measured variable,   the  Gini  coefficient  \cite{Gini1914},  is a classical indicator of non-uniform distributions.

The Theil index \cite{Theil1967} is defined by
\begin{equation}\label{Theilindexeq}
    Th =-\frac{1}{N}\sum_{i=1}^N \frac{z_i}{\sum_i z_i}  \; \ln\left( \frac{z_i}{\sum_i z_i} \right).
\end{equation}
It seems obvious that  the Theil index  can be expressed in terms of  the negative entropy
\begin{equation}\label{entropyeq}
   H = -\sum_{i=1}^N\frac{z_i}{\sum_i z_i} \; \ln \left(\frac{z_i}{\sum_i z_i}\right)
\end{equation}
indicating the deviation  from the maximum disorder entropy, $ln(N)$,
\begin{equation}\label{HTheq}
   H \;=\;ln (N) - Th \;\;or\;\; Th= ln(N) - H.
\end{equation}
  The exponential entropy \cite{Campbell66} is 
  \begin{equation} \label{expentropyeq}
 E=  exp(-H) = \Pi_{i=1}^N p_i^{p_i}.
 \end{equation}
 
The Hirschman-Herfindahl  index (HHI)  \cite{Hirschman}
  is an indicator of the "concentration" of variables, the
"amount of competition" between the months, here. The higher the value of HHI, the
smaller the number of months with a large value of (submitted, or accepted, or accepted if submitted) papers in a given month.
Formally, adapting the HHI notion to the present case, 
\begin{equation}\label{HHIeq}
HHI =  \sum_{i=1}^N \left(\frac{z_i}{\sum_i z_i} \right)^2.
\end{equation}
Notice that $HHI=\sum_{i=1}^N\; p_i^2$. 

The Gini coefficient $Gi$ \cite{Gini1914} has been widely used as a measure of  
    income \cite{AtkinsonBourguignon2014handbook} or wealth inequality \cite{CerquetiAusloosQuQuwealth,SSD15RCMAseanalysisIThagio}; nowadays,  it is widely used
 in  many  other  fields.  
In brief,  defining first  the Lorenz curve $L(r)$ as the percentage 
 contributed by the bottom $r$ of the variable population
to the total  value $\sum_r z_r$  of the measured (and  now  ranked)  variable $z_r$, i.e., 
$ p_r=[z_r/\sum_r z_r]$,
  one obtains the  Gini coefficient   as twice the area between this Lorenz curve and the diagonal line in the $[r, L(r)]$ plane;  such a diagonal  represents perfect equality; whence, $Gi=0$  corresponds to perfect equality of the $z_r$ variables.  
       

Having set up the framework and presented the definition of the indices to be calculated, we turn to the data and its analysis, in Section \ref{Data} and Section \ref{analysis} respectively. Their discussion and 
 comments on the present study, together with  a remark on  its limitations,   
  are found in the conclusion Section \ref{reasoning}.
   
\section{Data}\label{Data}

In order to develop the method  measuring the disorder of the time series, let us recall the necessary data.  The raw data can be found in  \cite{SCIM19MAONAD}.  For completeness, 
let the time series of submitted and of accepted papers if submitted during a given month to $JSCS$ and to $Entropy$ be recalled through   Fig. \ref{3yrtimeseries} for the  years  in which the full data is available, i.e. for which the final decisions have been made on the submitted papers. 

Let us introduce notations:

\begin{itemize} 
\item the number of monthly submissions  in a given month ($m= 1,\dots,12$) in   year ($y$)  is called $N_s^{(m,y)}$ 
\item the percentage of this set  is the probability of submission in a given month for     a  specific  year

$q_s ^{(m,y)}= N^{(m,y)}_s/\sum_m N^{(m,y)}_s$ 

\item similarly, one  can define $N_a^{(m,y)}$, as being   the number of accepted papers  when submitted  in   year ($y$)  in a specific  month ($m$),
\item and for the related percentage, one has $q_a^{(m,y)}= N^{(m,y)}_a/\sum_m N^{(m,y)}_a$;  

\item more importantly, for authors,   the (conditional) probability of a paper acceptance when submitted in a given month may be  considered and estimated  before submission
\begin{equation}
 p_{(a|s)}^{(m,y)}= N^{(m,y)}_a/ N^{(m,y)}_s 
\end{equation}
 \end{itemize}
 
 Thereafter, one can deduce the relevant "monthly information entropies"
 
 \begin{itemize} 
\item $S_s ^{(m,y)}=-q_s ^{(m,y)}\;  ln (q_s ^{(m,y)})$
\item $S_a^{(m,y)}=-q_a ^{(m,y)}\;  ln (q_a ^{(m,y)})$
\item $S_{(a|s)}^{(m,y)}=- p_{(a|s)}^{(m,y)} \; ln(p_{(a|s)}^{(m,y)})$
 \end{itemize}
 
 and the overall information entropy:
  \begin{itemize} 
\item $ S_s ^{(y)}= \sum_m S_s ^{(m,y)}$  
\item $ S_a^{(y)}= \sum_m S_a ^{(m,y)}$  
\item $S_{(a|s)}^{(y)} =\sum_m S_{(a|s)}^{(m,y)}$  
 \end{itemize}
 in order to pin point whether the yearly distributions are disordered.     
 
 Moreover, we can discuss the data not only comparing different years, but also the cumulated data per month in the examined time interval 
 as if all years are "equivalent" : \begin{itemize} 
\item
$C_s^{(m)}=\sum_y N_s^{(m,y)}$, from which one deduces  \item $q_s^{(m)} =  C_s^{(m)}\; /\; \sum_m C_s^{(m)}$
\item  and similarly for  the accepted papers  $C_a^{(m)}=\sum_y N_a^{(m,y)}$, and \item $q_a^{(m)} =  C_a^{(m)}\; /\; \sum_m C_a^{(m)}$
\item leading to the ratio between cumulated monthly data
\begin{equation}
  q_{(a|s)}^{(m)}=  C_a^{(m)}/C_s^{(m)},
  \end{equation}
 \item and  to  the corresponding "monthly cumulated entropy",
 $S_{(a|s)}^{(m)} = -  q_{(a|s)}^{(m)} \; ln(q_{(a|s)}^{(m)})$, 
 \item finally to  
  $S_{(a|s)}= \sum_m S_{(a|s)}^{(m)}$
   \end{itemize}
which will be called the "conditional entropy".

Relevant values are given in Tables 1-4 both for $JSCS$  and for $Entropy$.
The diversity and the inequality index values are given in Table 5. Most of the results stem for  the use of a free online software \cite{Ktauweb2014}. 
 
  \section{Data analysis}\label{analysis}
 
 \subsection{Data}
 
 First, notice that the 3 -year long time series in itself is not part of the main aim of the paper; this is  because we intend to compare data  with an equivalent number of  degrees of freedom,  i.e.  11, for all studied cases. Nevertheless, for completeness, and in order not to distract readers from our framework,  we provide relevant figures, but  in Appendix, together with a note on  the  corresponding discrete Fourier transform. 
 
 \subsection{Analysis  }
 
 The relevant values for the various indices,   given in Tables 1-4, both for $JSCS$  and for $Entropy$, serve for the following analysis.  We consider 3 aspects:
 (i) {\it a posteriori } features findings;, (ii) non-linear entropy indices, and (iii) forecasting aspects.
 
  \subsubsection{A posteriori  features findings }
 Browsing through Table 1, it can be noticed that the distribution of probabilities of submissions is weaker during  the February-May months for $JSCS$, but is rather high  for the fall and winter months.  For $Entropy$,  the highest probability of submissions also occurs  in October-December, and is preceded by a low rate of submissions, the lowest being in  February and in August, should one say at vacation times. Let us recall that the extremum entropy  (for "perfect disorder") is here $  ln(12) \simeq 2.4849$.
 
 Apparently this submission evolution pattern is  reflected, see Table 2,  in the acceptance rate, except for $JSCS$ which has a low acceptance rate for papers submitted in winter 2014. For $Entropy$, the weaker acceptance rate occur for papers submitted during August-September months, say end of summer time. 
 
 Statistical tests, e.g., $\chi^2$ , can be provided to ensure the validity of these findings for percentages, but taking into account the number of observations.   
 In all cases such a test demonstrates that the distributions are far from uniform, suggesting to look further for  the major deviations.  See a discussion of others texts in subsection \ref{forecasting}
 
 However,  $q_a^{(m,y)} $ values only measure the probability of monthly acceptances without considering the number of submissions in a given month. It is in this respect more appropriate to
  look  at  the conditional probabilities, $q_{(a|s)}^{(m)}$ , as  in Table 3.  For $JSCS$, the highest values of are found  for winter months:  $q_{(a|s)}^{(m)}$ has a notable maximum in January. and the lowest for spring-summer time, from March till August. There is a shift of such a pattern for $Entropy$: the highest conditional probabilities occur during spring time, except in 2016.
 
The  corresponding values of the monthly entropy,  for the given years and for the cumulated distributions, are found in Table 4. All values of the entropy are  remarkably $\simeq 4.1$, both for $JSCS$ and $Entropy$, suggesting some sort of universality.   One can notice that the entropy steadily increases as a function of time both for $JSCS$ and $Entropy$,  - the growth rate being about twice as large for the latter journal.  This is somewhat slightly surprising since one   should expect an averaging effect in the case of $Entropy$ because of the multidisciplinarity of involved topics. Comparing such values indicate that the distributions are far from uniform\footnote{The slight difference between the last lines of Table 3 and Table 4, displaying the "conditional entropy" is merely due to rounding errors.} indeed.
 
   \subsubsection{Non-linear entropy indices} 
  The diversity and  inequality   measures given in Table 5.  The diversity index $ ^1D $  is remarkably similar for both journals ($\sim 11$) for the submitted papers and accepted papers distributions. The similarity  holds also for the HHI $\simeq 0.087$, although a little bit lower for the $Entropy$ journal $\simeq 0.085$.  The diversity index for the conditional  probability distributions is however rather different:    both increase as a function of time, indicating an increase in concentrations for the in favor of relevant months. This increase rate  is much higher for $Entropy$ than for $JSCS$.
  
  The inequality between months is rather low, as further well  seen in the Gini coefficient; there is a weak inequality between months. However,  there is  a factor $\sim 2$ in favor of $JSCS$, which we interpret as due to the greater specificity of $JSCS$,  implying a  smaller involved community and   specially favored topics. This numerical observation  reinforces what can be deduced from  the Theil index, 
  whence inducing the same conclusion.
  
     \subsubsection{Forecasting aspects} \label{forecasting}
     Considering the rather small sizes of  both samples (not our faults!), it is of interest to discuss the significance of the findings, in some sense in view of suggesting some  "strategy" after the "diagnosis".
     The  notions of "false positives" and "false negatives", as in medical testing, can be applied in our framework.
     
      In brief, a "false positive" occurs as an error when    a test result improperly indicates the presence  (high probability) of an outcome,   when
in reality it is not present; obviously, {\it a contrario}  a "false negative" is an error in which a test result improperly indicates no presence of a condition (the result is negative), 
      when in reality it is present. This corresponds to  rejecting (or accepting) a null hypothesis, e.g.,  in econometrics.
    Thus, two statistical tests have been  used  for such a discussion: (i) the $t-$Student test and  (ii) the $z-$test.  Recall that they are used if either one does not know or one knows  the variance (or standard deviation)  of the sample and test distributions. Such characteristics are given in Table 1-4 for each relevant quantity.  
    
    For completeness, one has also given the confidence interval [$\mu-2\:\sigma\;,\; \mu + 2\; \sigma$]. It is easily seen that there is no outlier. This observation would lead, like other authors, to claim that there is no anomaly in the monthly numbers and subsequent percentages, in contradistinction with the $\chi^2$ values and tests.  We should here point out that the $t-$Student test leads to a $p$-value  $<$ 0.0001, whence to a quite significant result. Concentrating our 	 attention to the  (monthly and annual) conditional probabilities $N_a/N_s$, the $z-$ test gives the significance reported in Table 4. The values (so called $\alpha$, or {\it error of type I}) in  hypothesis testing, indicate that the correct conclusion is to reject the null hypothesis and to consider the existence of   "false positives". This is essentially due to the sample size. It is remarkable that the order of magnitude differs for $JSCS$ and for $Entropy$.
      
      
\section{  Conclusion}\label{reasoning}

  The  data   on the number of submitted papers is relevant for editors, and the more so nowadays for publishers due to the automatic handling of papers.  The relative number of accepted papers is less significant in that respect, but the conditional probability of having an accepted paper if it is submitted in a given month is much relevant for authors. 
 Authors expect fast and (hopefully) positive response from journals as they are probably interested to discover the best timing for their submission in order to avoid possible editor overload negative effect in a particular moment. For these authors,   the possible seasonal bias issue  is expected to be relevant, as they would like to know whether a specific month of submission will increase the chance that their paper will be accepted.
  Thus, the  probability of acceptance, the so called "acceptance rate" is the relevant variable to be studied! Instead of $\chi^2$ tests or observing the "confidence interval" on monthly  distributions, we  have proposed a new line of approach: considering the diversity and inequality in the  distributions  of papers submitted, accepted, or accepted if submitted in a given month through information indices, like the Shannon entropy \cite{crooks}, the diversity index, the Gini coefficients and the  Hirschman-Herfindahl     index. 
   
From this cases study, a seasonal bias  seems stronger in the specialized ($JSCS$) journal.  The features are  emphasized because we use a non linear transformation of the data, through information concepts, having their usefulness demonstrated in many other fields \cite{ClippeAusloosTheil}.  In the present cases, the seasonal  bias effects are observed. The overall significance and the universality features might have to be re-examined if more data was available. Indeed 
the $p-$values (so called $\alpha$, or {\it error of type I}) in  hypothesis testing, indicate that the correct conclusion is  to consider the existence of   "false positives".

Our  outlined findings  suggest  intrinsic behavioral hypotheses for future research.  Complementary aspects must be used as ingredients in order to understand whether some seasonal bias occurs \cite{Nedicefficiency,Nedicauthorsperspective}. 
  One has  markedly to take into account the scientific work environment, beside the journal favored topics.
   
   \vskip0.5cm
\section*{Acknowledgements}

 MA greatly thanks the MDPI  {\it Entropy} Editorial staff for  gathering and cleaning up the raw data, and in particular
Yuejiao Hu, Managing Editor.   Thanks go also to the reviewers and $Entropy$ editor. 
\clearpage

\begin{figure}
\includegraphics[height=16.8cm,width=16.8cm]{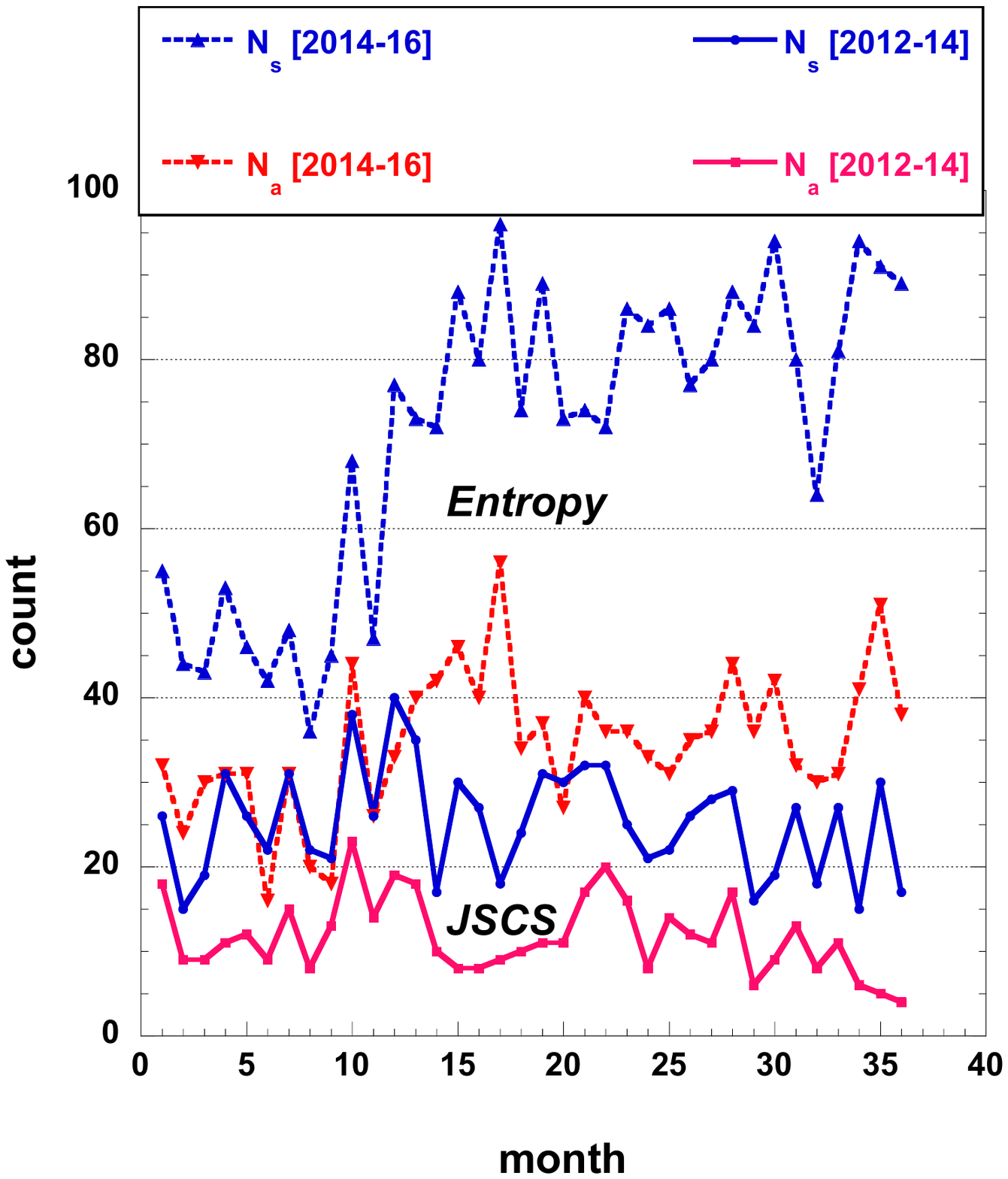} 
 \caption{ Number of papers submitted 
 and number of papers accepted if submitted 
during a given month,  to $JSCS$ and to $Entropy$, in  the examined 36 months of the 3-year time interval, [2012-2014] and [2014-2016], respectively.} \label{3yrtimeseries} 
\end{figure}

{\bf Appendix. Time series data}
\vskip0.5cm
The time series of submitted and of accepted papers if submitted during a given month to $JSCS$ and to $Entropy$ are given in Fig. \ref{3yrtimeseries}. The distribution are markedly non uniform. Nevertheless, with such rather short series, one can observe some periods more important than others. One can also observe that $Entropy$, a rather new journal, is  attracting more submission, since 2015, and having an increased rejection rate. Some "parallelism" in the numbers of submitted and accepted if submitted papers in a given month seems apparent for $JSCS$.

The two largest  amplitudes of frequency $f$  in $Month^{-1}$, or  (periods),  resulting from a Fourier analysis of the 3-year time series  for $N_s$  papers submitted or  $N_a$  accepted if submitted during a given month to $JSCS$ and $Entropy$ are given in Table \ref{amplitudesFourierfrequency}.  The year period is, in 3 cases, one of the two most important ones;  the trimester period is the most important for submitted papers to $JSCS$, and the next largest for $N_a$ to $JSCS$, indicating the more relevant timing for the journal, more prone toward academic authors than $Entropy$. 
\clearpage
{\bf Computational notes}

$https://www.medcalc.org/calc/test_-one_-mean.php$

This procedure calculates the difference of an observed mean with a hypothesized value. A significance value (P-value) and 95\% Confidence Interval (CI) of the observed mean is reported. The P-value is the probability of obtaining the observed mean in the sample if the null hypothesis value were the true value.

The P-value is calculated using the one sample $t$-test, with $t$ calculated as: 
\begin{equation}
t = \frac{\mu -k }{\sigma / \sqrt{N}}
\end{equation}
 
where the hypothesized mean is $k$ and the standard deviation  $\sigma$. In the present context the hypothesized mean corresponds to that of  the uniform distribution.
Recall that the P-value is the area of the $t$ distribution,  which for $N-1$ degrees of freedom, that falls outside $\pm \;t$. 
\clearpage

\clearpage 

\begin{table} 
 \caption{  Number of papers   $ N_s^{(y)}$ and monthly percentage  $q_s^{(m,y)}$ of papers submitted in a given year  ($y$) and month ($m$), respectively   to {\it JSCS} in  2012, 2013, and 2014,   and to $Entropy$ in 2014, 2015, and 2016;  $q_s^{(m)}$ is obtained after  summing the events of each year for a given month, i.e. from  $C_s^{(m)}$; last lines: $\chi^2$  and entropy; recall  that  $ln(12)\simeq$ 2.4849 and  $\chi^2_{11}(0.95\%)=4.5748$. } 
  \begin{center}
 \label{T1psubmittedJSCSENTROPY} 
      \begin{tabular}{|c||c|c|c|c||c|c|c|c|c|c|c|c|}
      \hline
   &      \multicolumn{4}{|c||}{{\it JSCS}} &      \multicolumn{4}{|c|}{{\it Entropy}} 	\\ \hline   
      \hline       
    $ N_s^{(y)}$&	317	&	322	&	274	&	913	&	604 	&	961 	&	1008 	&	2573 	\\ \hline \hline
&	$q_s^{(m,y)}$&$q_s^{(m,y)}$&	$q_s^{(m,y)}$ & $q_s^{(m)}$  &	$q_s^{(m,y)}$&$q_s^{(m,y)}$&	$q_s^{(m,y)}$ & $q_s^{(m)}$ \\ \hline
   $ y= $   & $ 2012 $ & $ 2013$  &	$ 2014$ &  [2012-14]\ &  $  2014 $ & $ 2015$  &	$ 2016$ & [2014-16]\ \\ \hline
Jan.	&	0.08202	&	0.10870	&	0.08029	&	0.09091	&	0.09106	&	0.07596	&	0.08532	&	0.08317	\\
Febr.	&	0.04732	&	0.05280	&	0.09489	&	0.06353	&	0.07285	&	0.07492	&	0.07639	&	0.07501	\\
Mar.	&	0.05994	&	0.09317	&	0.10219	&	0.08434	&	0.07119	&	0.09157	&	0.07937	&	0.08201	\\
Apr. 	&	0.09779	&	0.08385	&	0.10584	&	0.09529	&	0.08775	&	0.08325	&	0.08730	&	0.08589	\\
May	&	0.08202	&	0.05590	&	0.05839	&	0.06572	&	0.07616	&	0.09990	&	0.08333	&	0.08784	\\
June	&	0.06940	&	0.07453	&	0.06934	&	0.07119	&	0.06954	&	0.07700	&	0.09325	&	0.08162	\\
July	&	0.09779	&	0.09627	&	0.09854	&	0.09748	&	0.07947	&	0.09261	&	0.07937	&	0.08434	\\
Aug.	&	0.06940	&	0.09317	&	0.06569	&	0.07667	&	0.05960	&	0.07596	&	0.06349	&	0.06724	\\
Sept.	&	0.06625	&	0.09938	&	0.09854	&	0.08762	&	0.07450	&	0.07700	&	0.08036	&	0.07773	\\
Oct.	&	0.11987	&	0.09938	&	0.05474	&	0.09310	&	0.11258	&	0.07492	&	0.09325	&	0.09094	\\
Nov.	&	0.08202	&	0.07764	&	0.10949	&	0.08872	&	0.07781	&	0.08949	&	0.09028	&	0.08706	\\
Dec.	&	0.12618	&	0.06522	&	0.06204	&	0.08543	&	0.12748	&	0.08741	&	0.08829	&	0.09716	 \\ \hline
$\chi^2$ &	23.278	&	14.075	&	14.964	&	15.811	&	 29.497	 &	9.377  	&	9.333 	&	 20.236 	\\  \hline
entropy &2.4487 &2.4620 &2.4569 &2.4760 &2.4621 &2.4801 &2.4801 &2.4809  \\\hline \hline
  Mean	& 	0.08333	 & 	0.08333	 & 	0.08333	 & 	0.08333	 & 	0.08333	 & 	0.08333	 & 	0.08333 & 	0.08333 \\
 Std Dev	&	0.02359	 & 	0.01820	 & 	0.02034	 & 	0.01145	 & 	0.01923	 & 	0.00860	 & 	0.00837 & 	0.00772\\
 $\mu-2\sigma$	&0.03616 	& 0.04694 	& 0.04265 	& 0.06043 	& 0.04486 	& 	0.06614 	& 	0.06658  & 0.06790 \\
 $\mu+2\sigma$	& 0.13051 & 0.11973 & 0.12401 & 0.10624 & 0.12180 & 0.10053 & 0.10008 & 0.09877 \\\hline
 $t-stat$ & 654.12	 & 	854.49	 & 	705.30	 & 	2287.08 & 	1107.62&  3124.03 & 	3287.43 & 	5694.50 \\
$ signif.(p<)$ & 0.0001	 & 	0.0001	 & 	0.0001 & 	0.0001	 & 	0.0001	 & 	0.0001	 & 	0.0001 & 0.0001 \\  \hline
 \end{tabular}   \end{center}
  \end{table}   

\begin{table} 
 \caption{  Number of papers   $ N_a^{(y)}$ and monthly percentage  $q_a^{(m,y)}$ of papers accepted when submitted in a given year  ($y$) and month ($m$) respectively   to {\it JSCS} in  2012, 2013,  and 2014,  and to $Entropy$ in 2014, 2015, and 2016; $q_a^{(m)}$ is obtained after  summing the events of each year for a given month, i.e. from  $C_a^{(m)}$; last lines: $\chi^2$  and entropy; recall  $ln(12)\simeq$ 2.4849, and  $\chi^2_{11}(0.95\%)=4.5748$.} 
  \begin{center}
 \label{T2pacceptedJSCSENTROPY} 
      \begin{tabular}{|c||c|c|c|c||c|c|c|c|c|c|c|c|}
      \hline
   &      \multicolumn{4}{|c||}{{\it JSCS}} &      \multicolumn{4}{|c|}{{\it Entropy}} 	\\ \hline   
      \hline   
        $ N_a^{(y)}$&	160 	&	146 	&	116 	&	422 	&336 	&	467 	&	447 	&	1250 	\\ \hline \hline
&	$q_a^{(m,y)}$&$q_a^{(m,y)}$&	$q_a^{(m,y)}$ & $q_a^{(m)}$  &	$q_a^{(m,y)}$&$q_a^{(m,y)}$&	$q_a^{(m,y)}$ & $q_a^{(m)}$ \\ \hline
   $ y= $   & $ 2012 $ & $ 2013$  &	$ 2014$ &  [2012-14]\ &  $  2014 $ & $ 2015$  &	$ 2016$ & [2014-16]\ \\ \hline
Jan.	&	0.11250	&	0.12329	&	0.12069	&	0.11848	&	0.09524	&	0.08565	&	0.06935	&	0.08240	\\
Febr.	&	0.05625	&	0.06849	&	0.10345	&	0.07346	&	0.07143	&	0.08994	&	0.07830	&	0.08080	\\
Mar.	&	0.05625	&	0.05479	&	0.09483	&	0.06635	&	0.08929	&	0.09850	&	0.08054	&	0.08960	\\
Apr.	&	0.06875	&	0.05479	&	0.14655	&	0.08531	&	0.09226	&	0.08565	&	0.09843	&	0.09200	\\
May	&	0.07500	&	0.06164	&	0.05172	&	0.06398	&	0.09226	&	0.11991	&	0.08054	&	0.09840	\\
June	&	0.05625	&	0.06849	&	0.07759	&	0.06635	&	0.04762	&	0.07281	&	0.09396	&	0.07360	\\
July	&	0.09375	&	0.07534	&	0.11207	&	0.09242	&	0.09226	&	0.07923	&	0.07159	&	0.08000	\\
Aug. 	&	0.05000	&	0.07534	&	0.06897	&	0.06398	&	0.05952	&	0.05782	&	0.06711	&	0.06160	\\
Sept.&	0.08125	&	0.11644	&	0.09483	&	0.09716	&	0.05357	&	0.08565	&	0.06935	&	0.07120	\\
Oct.	&	0.14375	&	0.13699	&	0.05172	&	0.11611	&	0.13095	&	0.07709	&	0.09172	&	0.09680	\\
Nov. 	&	0.08750	&	0.10959	&	0.04310	&	0.08294	&	0.07738	&	0.07709	&	0.11409	&	0.09040	\\
Dec. 	&	0.11875	&	0.05479	&	0.03448	&	0.07346	&	0.09821	&	0.07066	&	0.08501	&	0.08320	 \\ \hline
$\chi^2 $& 18.200	&17.068	&18.276	&20.806	& 	23.4286	&14.8243	&11.7651 	&19.5802\\ 
  \hline
entropy	&2.4305 &2.4291 &2.4042 &2.4612 &2.4496 &2.4695 &2.4722 &2.4769\\  \hline
 \hline
  Mean	& 0.08333	 & 	0.08333	 & 	0.08333	 & 	0.08333	 & 	0.08333	 & 	0.08333	 & 	0.08333 & 	0.08333 \\
Std Dev	&	0.02935  & 0.02976  & 0.03455 & 0.01933 & 0.02298  & 0.01551 & 0.01412 & 0.01089 	\\
$\mu-2\sigma$ & 0.02462  & 0.02381  & 0.01424 & 0.04468 & 0.03737 & 0.05232 & 0.05509 & 0.06155 \\
 $\mu+2\sigma$	&0.14204 & 0.14285 & 0.15243 & 0.12199 & 0.12930 & 0.11435 & 0.11157 & 0.10512  \\\hline
$t-stat.$ &373.51 & 351.88 & 270.17 & 921.04 & 691.19 & 1207.53 & 1297.69 & 2813.71\\
$ signf.(p<)$ & 0.0001	 & 	0.0001	 & 	0.0001 & 	0.0001	 & 	0.0001	 & 	0.0001	 & 	0.0001 & 0.0001 \\  \hline
 \end{tabular}   \end{center}
  \end{table}

 \begin{table} 
  \caption{ Conditional   probability  $p_{(a|s)}^{(m,y)}  =  N_a^{(m,y)} /N_s^{(m,y)}$  of   having a   \underline{paper accepted}   \underline{if submitted in a given month} ($m$)   to {\it JSCS} or to {\it Entropy} in  a given year ($y$), and the corresponding cumulated conditional probability $q_{(a|s)}^{(m)}=  C_a^{(m)}/C_s^{(m)} = $ $ \sum_y N_a^{(m,y)}/ \sum_y N_s^{(m,y)}$;  the sum of such probabilities is given; on the last line is the here so called "conditional   entropy" ( $c.entr.$), either  $S_{(a|s)}^{(y)}$ or  $S_{(a|s)}$ .  }    \label{JSCSEntropypasmonth}
   \begin{center}
   \begin{tabular}{|c||c|c|c|c||c|c|c|c| }
      \hline    
    $month$   &   \multicolumn{4}{|c|}{{\it JSCS}} &      \multicolumn{4}{|c|}{{\it Entropy}} 	\\ \hline  
    &   \multicolumn{3}{|c|}{{\it $p_{(a|s)}^{(m,y)}$}} &     $q_{(a|s)}^{(m)}$& \multicolumn{3}{|c|}{{\it $p_{(a|s)}^{(m,y)}$}} 	& $q_{(a|s)}^{(m)}$\\ \hline  
   &   $2012$&    $2013$&$2014$&[2012-14]&$2014$&$2015$&$2016$&[2014-16]\\ \hline      
Jan. 		&	0.6923	&	0.5143	&	0.6364	&	0.6024	&	0.5818	&	0.5479	&	0.3605	&	0.4813	\\
Febr.		&	0.6000	&	0.5882	&	0.4615	&	0.5345	&	0.5455	&	0.5833	&	0.4545	&	0.5233	\\
March 	&	0.4737	&	0.2667	&	0.3929	&	0.3636	&	0.6977	&	0.5227	&	0.4500	&	0.5308	\\
April		&	0.3548	&	0.2963	&	0.5862	&	0.4138	&	0.5849	&	0.5000	&	0.5000	&	0.5204	\\
May		&	0.4615	&	0.5000	&	0.3750	&	0.4500	&	0.6739	&	0.5833	&	0.4286	&	0.5442	\\
June		&	0.4091	&	0.4167	&	0.4737	&	0.4308	&	0.3810	&	0.4595	&	0.4468	&	0.4381	\\
July		&	0.4839	&	0.3548	&	0.4815	&	0.4382	&	0.6458	&	0.4157	&	0.4000	&	0.4608	\\
Aug. 		&	0.3636	&	0.3667	&	0.4444	&	0.3857	&	0.5556	&	0.3699	&	0.4687	&	0.4451	\\
Sept. 	&	0.6190	&	0.5312	&	0.4074	&	0.5125	&	0.4000	&	0.5405	&	0.3827	&	0.4450	\\
Oct. 		&	0.6053	&	0.6250	&	0.4000	&	0.5765	&	0.6471	&	0.5000	&	0.4362	&	0.5171	\\
Nov. 		&	0.5385	&	0.6400	&	0.1667	&	0.4321	&	0.5532	&	0.4186	&	0.5604	&	0.5045	\\
Dec. 		&	0.4750	&	0.3810	&	0.2353	&	0.3974	&	0.4286	&	0.3929	&	0.4270	&	0.4160	\\
\hline \hline
$c.entr.$ &	4.0120 	&	4.0970 	&	4.1301 	&	4.2136 	&	3.7919 	&	4.1450 	&	4.2943 	&	4.1883\\ 
 \hline \hline 
sum	& 	6.0767	&	5.4809	&	5.0610	&	5.5375	&	6.6951	&	5.8343	&	5.3154	&	5.8266	\\ 
Mean ($\mu$)	&	0.5064	&	0.4567	&	0.4217	&	0.4615	&	0.5579	&	0.4862	&	0.4429	&	0.4856	\\
Std Dev &	0.1063	&	0.1271	&	0.1297	&	0.0770	&	0.1058	&	0.0737	&	0.0528	&	0.0432	\\
 $\mu-2\sigma$	&	0.2939 & 0.2026 & 0.1624 & 0.3075 & 0.3463 & 0.3387 & 0.3373 & 0.3992 \\
 $\mu+2\sigma$	&	0.7189  & 0.7109  & 0.6811  & 0.6154 	 & 0.7695  & 0.6337  & 0.5486  & 0.5719   \\\hline
$t-test$ &52.786 & 46.897 & 43.870 & 130.33& 67.933 & 135.995& 203.05 & 380.07    \\
$z-test$ &0.803 & 40.758 & 0.673 &  1.268 & 1.198 &  1.347& 1.291 & 2.190   \\
  $ p-level$ & 0.4221	 & 	0.4484	 & 	0.5012 & 	0.2047	 & 	0.2309	 & 	0.1780	 & 	0.1968 & 0.0285 \\  \hline
 \end{tabular}   \end{center}
  \end{table}

\begin{table} 
  \caption{  Monthly information Entropy and (last line) overall information entropy for specific years $S_{(a|s)}^{(m,y)}$ and for the cumulated data  over the relevant time interval $S_{(a|s)}^{(m)}$  for either journal so investigated; on the last line is the here so called "conditional   entropy", $c.entr.$,  either  $S_{(a|s)}^{(y)}$ or  $S_{(a|s)}$.}    \label{MonthlyinformationEntropy}
   \begin{center}
   \begin{tabular}{|c||c|c|c|c||c|c|c|c| }
      \hline    
    $month$   &   \multicolumn{4}{|c|}{{\it JSCS}} &      \multicolumn{4}{|c|}{{\it Entropy}} 	\\ \hline  
    & \multicolumn{3}{|c|}{{\it $S_{(a|s)}^{(m,y)}$}} & $S_{(a|s)}^{(m)}$& \multicolumn{3}{|c|}{{\it $S_{(a|s)}^{(m,y)}$}} & $S_{(a|s)}^{(m)}$\\ \hline  
   		& $2012$	&    	$2013$	&	$2014$	&	[2012-14]&	$2014$	&	$2015$	&	$2016$	&	[2014-16]\\ \hline      
January		& 0.25458	&	0.34199	&	0.28763	&	0.30531	&	0.31511	&	0.32963	&	0.36780	&	0.35196	\\
February		& 0.30650	&	0.31213	&	0.35686	&	0.33483	&	0.33062	&	0.31441	&	0.35839	&	0.33888	\\
March		& 0.35394	&	0.35247	&	0.36705	&	0.36785	&	0.25116	&	0.33909	&	0.35933	&	0.33619	\\
April			& 0.36765	&	0.36041	&	0.31308	&	0.36513	&	0.31369	&	0.34657	&	0.34657	&	0.33992	\\
May			& 0.35686	&	0.34657	&	0.36781	&	0.35933	&	0.26596	&	0.31441	&	0.36313	&	0.33109	\\
June			& 0.36565	&	0.36478	&	0.35394	&	0.36279	&	0.36765	&	0.35732	&	0.35996	&	0.36157	\\
July			& 0.35126	&	0.36765	&	0.35191	&	0.36155	&	0.28237	&	0.36489	&	0.36652	&	0.35702	\\
August		& 0.36785	&	0.36788	&	0.36041	&	0.36745	&	0.32655	&	0.36787	&	0.35517	&	0.36029	\\
September	& 0.29688	&	0.33603	&	0.36583	&	0.34258	&	0.36652	&	0.33253	&	0.36758	&	0.36031	\\
October		& 0.30390	&	0.29375	&	0.36652	&	0.31754	&	0.28168	&	0.34657	&	0.36190	&	0.34104	\\
November		& 0.33333	&	0.28562	&	0.29863	&	0.36257	&	0.32752	&	0.36453	&	0.32451	&	0.34518	\\
December	 	& 0.35361	&	0.36765	&	0.34045	&	0.36672	&	0.36313	&	0.36705	&	0.36337	&	0.36486	\\  \hline
 \hline 
 $c.entr.$	& 4.0120	&	4.0969	&	4.1301	&	4.2137	&	3.7919	&	4.1449	&	4.2942	&	4.1883\\
\hline \hline
 Mean	& 0.33433 & 0.34141 & 0.34418 & 0.35114  &  0.3160 &  0.34541 & 0.35785 &   0.34903 \\
 Std Dev &	0.03597 & 0.02924 & 0.02842 & 0.02131 & 0.03922 & 0.01963 & 0.01205 & 0.01162  \\
$\mu-2\sigma$	&	0.26240 & 0.28294 & 0.28734 & 0.30852 & 0.23755 & 0.30615 & 0.33376 & 0.32578 \\
$\mu+2\sigma$	&	0.40627 & 0.39989 &  0.40101 & 0.39375  & 0.39444 & 0.38467 & 0.38195 & 0.37227  \\\hline
 $t-stat.$ & 216.505 & 251.492 & 229.588 & 577.295 & 295.560 & 665.578 & 1060.13 & 1828.53   \\
  $ signf.(p<)$ & 0.0001	 & 	0.0001	 & 	0.0001 & 	0.0001	 & 	0.0001	 & 	0.0001	 & 	0.0001 & 0.0001 \\  \hline
\end{tabular}   \end{center}
  \end{table}   
  
            \begin{table} 
  \caption{  Diversity index,  the exponential entropy ($e.entr.$), Theil index,  Hirschman-Herfindahl  index, and Gini coefficient,   for specific years and for the cumulated data over the relevant time interval for the submitted, accepted, and accepted if submitted papers, respectively, to both investigated journals}    \label{diversityGiniThHHI}
   \begin{center}
   \begin{tabular}{|c||c|c|c|c||c|c|c|c| }
      \hline    
   $ $   &   \multicolumn{4}{|c||}{{\it JSCS}} &      \multicolumn{4}{|c|}{{\it Entropy}} 	\\ \hline  
$index$	& 	$2012$	&    	$2013$	&	$2014$	&	[2012-14]	&	$2014$	&	$2015$	&	$2016$	&	[2014-16]\\ \hline     \hline
 	& 	  \multicolumn{8}{|c|}{{\it submitted papers}}	 \\ \hline      
$^1$D	& 	11.574 	&	11.729 	&	11.669 	&	11.893 	&	11.730 	&	11.942 	&	11.943 	&	11.952	\\

$e.entr.$ &	0.08640	&	0.08526 &	0.08570 &	0.08408	&	0.08526 &	0.08373 &	0.08373 &	0.08367	\\
$Th$		&	0.03619	&	0.02287	&	0.02797	&	0.00893	&	0.02280	&	0.00480 	&	0.00480	&	0.00399	\\
HHI		&	0.08945	&	0.08698	&	0.08788	&	0.08478	&	0.08740	&	0.08415	&	0.08410	&	0.08399	\\
$Gi$		&	0.15063	&	0.11749	&	0.13139	&	0.07329	&	0.11369	&	0.05402	&	0.05192	&	0.04861	\\
 \hline  \hline
& 	  \multicolumn{8}{|c|}{{\it  accepted papers}}	 \\ \hline     
$^1$D	& 	11.364 	&	11.349 	&	11.069 	&	11.719 	&	11.584 	&	11.817 	&	11.848 	&	11.904	\\
$e.entr.$ &	0.08799 &	0.088114 & 0.09034 &0.08533 &0.08633 &0.08463 &0.08440 &	0.08401	\\
$Th$ 	&	0.05446	&	0.05578	&	0.08073	&	0.02371	&	0.03528	&	0.01539	&	0.01275	&	0.00803	\\
HHI		&	0.09281	&	0.09308	&	0.09646	&	0.08746	&	0.08914	&	0.08598	&	0.08553	&	0.08464	\\
$Gi$		&	0.18646	&	0.18949	&	0.22557	&	0.12164	&	0.14335	&	0.09404	&	0.08930	&	0.07027	\\
  \hline  \hline
& 	  \multicolumn{8}{|c|}{{\it  accepted papers if submitted in a given month}}	 \\ \hline     
$^1$D	& 	55.257 	&	60.158 	&	62.186 	&	67.602 		&	44.341 	&	63.116 	&	73.278 	&	65.912	\\
$e.entr.$ &	0.08504	&0.08634  &0.08737 &0.08438 &	0.08478 &	 0.08423  &0.08387 &	0.08364\\
$Th$		&	0.02022	&	0.03614	&	0.04727	&	0.01244		&	0.01716	&	0.01070	&	0.00641	&	0.00365	\\
HHI		&	0.08670	&	0.08924	&	0.09056	&	0.08546		&	0.08608	&	0.08509	&	0.08442	&	0.08394	\\
$Gi$		&	0.11355	&	0.15211	&	0.15965	&	0.08820		&	0.10083	&	0.08264	&	0.06189	&	0.04808	\\
\hline
\end{tabular}   \end{center}
  \end{table}

\begin{table} 
 \caption{ The two largest  amplitudes of frequency  $f$ in $Month^{-1}$, or  (periods),  resulting  from a Fourier analysis of the 3-year time series  for papers $N_s$   submitted or $N_a$   accepted if submitted during a given month to $JSCS$ and $Entropy$, as displayed in Fig. \ref{3yrtimeseries}.}  
  \begin{center}
 \label{amplitudesFourierfrequency} 
      \begin{tabular}{|c||c|c|c|c||c|c|c|c|c|c|c|c|}
     \hline
         &      \multicolumn{4}{|c||}{{\it JSCS}} &      \multicolumn{4}{|c|}{{\it Entropy}} 	\\ \hline 
         & $N_s$& $f$ & $N_a$ & $f$ & $N_s$ & $f$  &$N_a$&f   $f$  \\\hline
      \hline
1	&	125.42	&	0.3333 (3)	&	66.83 	&	0.0833 (12)	&	720.23 	&	0.0278 (36)	&	169.36&	0.0556 (18)\\
2	&	94.94 	&	0.3889 (2.57)	&	51.11	&	0.3333 (3)	&	378.38 	&	0.0833 (12)	&	164.15 	&	0.0833 (12) \\ 
 \hline
 \end{tabular}   \end{center}
  \end{table}  
  
   \end{document}